\documentclass[superscriptaddress,twocolumn]{revtex4-2}
\usepackage{graphicx}
\usepackage{bm}
\usepackage{dsfont}
\usepackage{amsmath, amsthm, amssymb,mathrsfs}
\usepackage{setspace}
\usepackage[utf8]{inputenc}
\usepackage[toc,page]{appendix}
\usepackage{epstopdf}
\usepackage{hyperref}
\usepackage{mdframed}
\usepackage{xcolor}
\usepackage{soul}

\def\beq{\begin{equation}}
\def\eeq{\end{equation}}
\def\beqn{\begin{eqnarray}}
\def\eeqn{\end{eqnarray}}

\makeatletter
\newcommand\be{\@ifstar{\[}{\begin{equation}}}
\newcommand\ee{\@ifstar{\]}{\end{equation}}}
\makeatother
\newcommand\bp{\begin{pmatrix}}
\newcommand\ep{\end{pmatrix}}

\newcommand{\ket}[1]{\left| #1 \right\rangle}

\begin{document}

\title{Properties of Invariant Set Theory}
\author{J.R.\ Hance}
\affiliation{Quantum Engineering Technology Laboratories, Department of Electrical and Electronic Engineering, University of Bristol, Woodland Road, Bristol, BS8 1US, UK}
\author{S.\ Hossenfelder}
\affiliation{Frankfurt Institute for Advanced Studies, Ruth-Moufang-Str. 1, D-60438 Frankfurt am Main, Germany}
\author{T.N.\ Palmer}
\email{tim.palmer@physics.ox.ac.uk}
\affiliation{ Department of Physics, University of Oxford, UK}
\date{\today}

\begin{abstract}
In a recent paper (\href{http://arxiv.org/abs/2107.04761}{arXiv:2107.04761}), Sen critiques a superdeterministic model of quantum physics, Invariant Set Theory, proposed by one of the authors. He concludes that superdeterminism is `unlikely to solve the puzzle posed by the Bell correlations'. He also claims that the model is neither local nor $\psi$-epistemic. We here detail multiple inaccuracies with Sen's arguments - notably that the hidden-variable model of quantum physics he uses to critique Invariant Set Theory bares no relation to Invariant Set Theory - and use this opportunity to lay out the properties of Invariant Set Theory as clearly as possible.
\end{abstract}

\maketitle

\section{Introduction}
\label{intro} 

Motivated by the fractal geometry of chaos, Invariant Set Theory \cite{Palmer2020Discretization} is a locally causal deterministic model of quantum physics. As outlined below, this model provides the rationale for a particular discretisation of complex Hilbert Space, which allows an ensemble interpretation of Hilbert states. Invariant Set Theory is based on the premise that the universe is a dynamical system evolving on a fractal set of measure zero in state space. This set is assumed to be invariant under the action of an as-yet unknown nonlinear deterministic dynamical law - hence the name Invariant Set Theory, hereafter {\sc IST}. The fractal geometry can be thought of as providing a `timeless' algebraic expression for the dynamical law. 

The essential assumption of {\sc IST} is that the laws of physics at the most basic level describe the geometry of this invariant set. This means that putative (mentally constructed, and hence conceivable) states which do not lie on the invariant set are, by construction, inconsistent with the laws of physics. Because of the measure-zero property of the invariant set, counterfactual states associated with worlds where some experiment might have been performed but wasn't, can typically lie off the invariant set. 

As discussed below, this property of counterfactual states provides a generic explanation in {\sc IST} of the quantum mechanical notion of complementarity, whereby quantum systems cannot be simultaneously be measured, using non-commuting observables, with well defined measurement outcomes. As discussed, this same property also provides a deterministic causal explanation of Bell-type experiments.For this reason  formally violates the assumption of Statistical Independence in Bell's Theorem, and the theory can be said to be superdeterministic \mbox{\cite{Hossenfelder2020Rethinking,Hossenfelder2020SuperdeterminismGuide}}.

The purpose of this paper is to provide a robust critique of a paper \cite{Sen2021Analysis} by Indrajit Sen, which itself critiques {\sc IST}. 
Sen's fundamental premise is that because {\sc IST} is not formulated precisely enough to be critiqued, he develops his own hidden-variable theory which he claims is consistent with the basic premises of {\sc IST}. He then shows that his hidden-variable theory does not have the properties claimed of {\sc IST}.

We argue here that even though the global geometric equations of the invariant set are unknown, its local structure in state space is indeed formulated precisely enough to analyse such matters as Bell's Theorem. Importantly, the definition of hidden variables, as labelling trajectories on the invariant set, is clear. For reasons which are unclear to us, as discussed in Section \mbox{\ref{counter}}, Sen chooses to develop a hidden-variable theory which bears no relation to {\sc IST}. In particular, Sen's assumption that experimenters have some control over aspects of the hidden variables of quantum systems has no correspondence in {\sc IST} (and indeed makes no sense to us). As a result, {\sc IST}'s precisely defined number-theoretic properties play no role in constraining the consistency of key counterfactual quantum experiments in Sen's hidden-variable model.

We conclude that the hidden-variable model developed by Sen is something completely separate to and independent of {\sc IST}. Hence none of Sen's conclusions are relevant to or apply to {\sc IST}. We continue to insist that IST provides a plausible deterministic locally causal interpretation of the violation of Bell's inequality, associated a violation of Statistical Independence. By construction, contrary to Sen's claim, states are ontic in {\sc IST} and quantum uncertainty epistemic.

\section{Discretisation of Hilbert Space}

The birth of quantum theory arose from Planck's bold suggestion that the energy states of light do not vary continuously, but in discrete jumps. Despite this insight, the state space of quantum theory, complex Hilbert Space, is itself continuous. In {\sc IST} - applying Planck's insight once more - this continuous space is replaced by a discretised space of Hilbert States.

It is crucially important to note that the {\sc IST}'s properties do not arise from \emph{any} discretisation of Hilbert Space. Instead they arise from a specific type of discretisation where complex phases are rational angles and the modulus squared of complex amplitudes are rational numbers. We can express this notion in finite terms by introducing a natural number $p$, assumed large, and requiring that complex phases are multiples $m/p$ of $2\pi$, and the modulus squared of complex amplitudes are of the form $n/p$, for integers $m$ and $n$. The key properties of {\sc IST} hold no matter how large is $p$. The essential number-theoretic result which {\sc IST} makes extensive use of is Niven's Theorem: that rational angles typically do not have rational cosines.

We have reviewed these facts to counter Sen's claim that IST is not precisely formulated. For the purposes of our analysis of quantum complementarity and Bell's theorem, it is.

\section {Use of Counterfactuals}
\label{counter}

Sen critiques {\sc IST} by constructing his own hidden variable version of {\sc IST}. However, this version departs so radically from {\sc IST} that no conclusions drawn from Sen's model have relevance to {\sc IST}.

The origin of the departure of Sen's model from {\sc IST} lies in the notion of a hidden variable. Here we follow Bell's original prescription - we assume that the spin value of a particle can be written as $S(\lambda, X)$ where $\lambda$ denotes the particle's so-called hidden variables and $X$ denotes measurement settings. This form makes explicit the notion that the hidden variables and the measurement settings are to be considered conceptually independent. We assume that experimenters are free to choose $X$. That is to say, they can base their choice of $X$ on the date of their grandmother's birthday, on the precise wavelength of light from a distant quasar, or from one of a million other whimsical factors. By contrast, in {\sc IST}, the hidden variables simply are what they are. In {\sc IST} it makes no sense, as Sen's hidden variable model seeks to do, to make some aspect of the hidden variables dependent on, and therefore controlled by, experimenter choices.

This is relevant when considering the notion of a quantum counterfactual, central to IST. Consider something simple, like a Mach-Zehnder interferometer experiment. Here we can imagine that $X=1$ corresponds to an apparatus setting where both half-silvered mirrors are in place and the apparatus performs an interferometric measurement. By contrast, $X=0$ corresponds to an apparatus setting where the half-silvered mirror is removed and the apparatus performs a `which-way' measurement.

Suppose in reality an $X=1$ experiment is performed. In IST, a counterfactual experiment on the \emph{same particle} corresponds to an $X=0$ experiment, where the particle's hidden variables are held fixed. What are these hidden variables and what does it mean to keep the hidden variables fixed?

At its most basic, $\lambda$ is simply a partial label for one of the trajectories on the invariant set. In terms of space-time, we can imagine $\lambda$ as describing the relationship between the particle described by $\lambda$ with other particles in the universe. Even in Newtonian mechanics, this is not a completely unusual notion. To ascertain the motion of a free particle in a given frame of reference, one needs to know whether that frame is inertial or not. This can be determined by the motion of that frame relative to the distant stars. In this sense the motion of the particle, and hence its properties, depends on its relationship to the distant mass of the universe - we call this Mach's Principle (a guiding concept for Einstein in formulating general relativity). However, there is nothing nonlocal about Mach's Principle. It does not imply correlations between causally disconnected physical events in space time. One should not conflate a global constraint with nonlocality.

The crucial question when we consider counterfactual experiments such as the one above, is whether it is consistent with the underlying theory to vary $X$ keeping the relationship of the quantum particle described by $\lambda$ with the other particles in the universe fixed.

Hence, in thinking physically what is this counterfactual world, we conclude it is one where everything in the universe, except for the variable which defines the measurement settings, is held fixed relative to the quantum particle. Hence, for example, the moons of Jupiter are held fixed, whilst the second half-silvered mirror of the Mach-Zehnder interferometer is removed.

It is worth noting that Bell himself realised that the theoretical consistency of such counterfactuals was \emph{critical} to the interpretation of his eponymous theorem. He wrote
\cite{Bell2004Speakable}
\begin{quote}
``I would insist here on the distinction between analysing various physical theories, on the one hand, and philosophising about the unique real world on the other hand. In this matter of causality, it is a great inconvenience that the real world is given to us once only. We cannot know what would have happened if something had been different. We cannot repeat an experiment changing just one variable the hand of the clock will have moved, and the moons of Jupiter. Physical theories are more amenable in this respect. We can calculate the consequences of changing free elements in the theory, be they only initial conditions, and so can explore the causal structure of the theory. I insist that [Bell's Theorem] is primarily an analysis of certain kinds of physical theory. ''
\end{quote}
{\sc IST} is a `certain kind of physical theory' where varying $X$ but keeping the moons of Jupiter fixed relative to the quantum particle being measured, is inconsistent with the presumed laws of physics. The reason for this is a subtle one. If the relevant Hilbert state (on the discretised Bloch Sphere) for an interferometric measurement $X=1$ is
\begin{equation}
\cos \frac{\phi}{2} |0\rangle + \sin \frac{\phi}{2} |1\rangle    
\end{equation}
then the Hilbert state (on the discretised Bloch Sphere) for the corresponding which-way measurement $X=0$ is
\begin{equation}
\frac{1}{\sqrt 2}(|1\rangle +ie^{i \phi} |0\rangle
\end{equation}

We now invoke Niven's theorem: if $X=1$ then the phase angle $\phi$ must have a rational cosine. Hence, typically, $\phi$ itself cannot be a rational angle and hence the Hilbert State associated with $X=0$ is typically undefined in IST.

In his analysis of IST, Sen introduces a different kind of hidden-variable theory, one where hidden variables are explicitly linked to measurement settings. He then classifies hidden variables into those under the control of the experimenter and those (associated with uncontrollable fluctuations in experimental settings) that aren't. Why he should want to do this is not clear to us. However, it means that `keeping the hidden variables fixed' has a completely different meaning in Sen's hidden variable model to IST. In particular, Sen allows the counterfactual state to differ from the real-world state by these uncontrollable fluctuations.

It is easy to see why this completely negates the properties of IST. Suppose, after an experimenter has performed an interferometric measurement on a particle, she repeats the experiment on the same particle but where the second half-silvered mirror has been removed. Clearly that is a permissible sequence of experiments. IST allows this sequence precisely because the moons of Jupiter have moved between the first and second experiments. That is to say, the relationship between the quantum particle and the rest of the universe has changed. Put another way, the existence of uncontrollable fluctuations makes the hidden variable different in the second real-world experiment than in the first counterfactual experiment. By not including these `uncontrollable fluctuations' in the specification of the particle's hidden variable, Sen has totally deviated from {\sc IST}. His critiques of {\sc IST} are, therefore, completely nullified by dropping this essential feature of $\sc IST$.

Perhaps it might be argued that making a distinction between rational and irrational angles and in so doing making the hidden variables critically dependent on the `uncontrollable fluctuations', {\sc IST} is a very finely-tuned theory. However, this is not so. Points which lie off the invariant set are in a $p$-adic sense distant from points which lie on the invariant set, no matter how close they may seem from a Euclidean perspective. Indeed, the $p$-adic metric is more relevant than the Euclidean metric when considering fractal geometries (the set of $p$-adic integers is homeomorphic to a Cantor Set with $p$ iterated pieces). It is critically important - and typically overlooked - when accusing a theory of being fine tuned, to state with respect to which metric or measure the tuning is fine. With respect to the natural ($p$-adic) metric or (Haar) measure associated with fractal geometry, {\sc IST} is not finely tuned at all.

A second example where Sen's model obscures the properties of {\sc IST} is the sequential Stern-Gerlach (SG) experiment. Here the spin of a particle is prepared by SG1, and then passes through SG2 and SG3. Using Niven's theorem again, a counterfactual experiment where we keep the particle hidden variable fixed, but swap SG2 and SG3, lies off the invariant set. Sen does not accept this result. He writes:
\begin{quote}
``Here we make use of the fact that the exact apparatus orientations are continuously fluctuating with time (see section II). If the ordering is changed, then the apparatuses will get used at different times than previously. Therefore, the exact apparatus orientations will also change.''
\end{quote}
Again, this completely confuses the meaning of a counterfactual in IST (and indeed in quantum theory). Whatever the nominal orientations of SG2 and SG3 as far as the experimenter is concerned, there is a definite precise orientation of the magnetic field in the real world, replete with uncontrollable fluctuations. The counterfactual experiment being considered is one where we replace the \emph{exact} orientation of SG2 with the \emph{exact} orientation of SG3. IST shows why - using number theory - it is impossible for the particle to be measured simultaneously in the \emph{exact} SG2 and SG3 directions.

By not engaging this vital number-theoretic incompatibility, Sen makes another mistake. In considering the sequential Stern-Gerlach experiment, Sen assumes that in {\sc IST} it is necessary for the individual Stern-Gerlach devices to be precisely orthogonal. This is not the case at all! The relative orientation of the devices is effectively arbitrary within the constraints of the rationality assumption. The number-theoretic incompatibility that lies at the heart of {\sc IST} does not assume orthogonal devices. 

 \section{Sen's Claim that IST is Non-Local}
 
In the CHSH version of Bell's Theorem we can write $S=S(\lambda, X,Y)$ for the spin of Alice's particle, where $X$ and $Y$ denote Alice and Bob's measurement settings respectively. In a local theory we write $S=S(\lambda,X)$ for Alice's particle (and $S=S(\lambda, Y)$ for Bob's particle).
 
Using Niven's theorem once more, it can be shown in IST that if $\{\lambda, X,Y\}$ denotes a CHSH experiment in the real world and hence lying on the invariant set, then $\{\lambda, X, Y'\}$  denotes a counterfactual CHSH experiment which does not lie on the invariant set. Here $Y'=0$ when $Y=1$ and \emph{vice versa}. In this way, the value of $Y$ in $S(\lambda, X,Y)$ for Alice's particle is redundant, given $\lambda$ and $X$. Similarly arguments can be made for Bob's particle. In this sense IST is locally causal.

If we introduce a probability measure on the space of hidden variables, then the above property of IST can be expressed as follows:
\begin{equation}
    \rho(\lambda |X Y) \ne 0 \implies \rho(\lambda | X Y')=0
\end{equation}
This makes IST superdeterministic (though we now prefer to use the word `supermeasured' \mbox{\cite{Hance2021StatInd}}) which is to say that the Statistical Independence assumption $\rho(\lambda |XY)=\rho(\lambda)$ is violated.  In a superdeterministic theory, one cannot assume that $\lambda$ can be varied independently of $X$ and $Y$.  One can put it like this: no matter how Alice and Bob choose their measurement settings (e.g., based on dates of grandmothers birthdays, or wavelength of light from distant quasars), the cosine of the angular distance between Alice and Bob's \emph{exact} measurement orientations can always satisfy IST's rationality conditions. However, given this, then by Niven's theorem it is impossible to satisfy IST's rationality condition for the cosine of the angular distance between Alice's actual \emph{exact} measurement orientation and Bob's \emph{exact} counterfactual measurement orientation.

In this sense, the full specification of the prepared state implicitly contains information about crucial aspects of the measurement settings. Indeed, it is a common mistake to not realize that the measurement settings effectively appear twice in any superdeterministic theory, once in the information that determines the evolution of the prepared state, and once as the setting itself. Needless to say, if one does not keep in mind that the measurement setting cannot be varied independently of the hidden variables, one arrives at the odd conclusion that the superdeterministic theory violates local causality. But of course, the entire reason Bell considered superdeterminism was that it's a way to restore local causality. Importantly, the definition of hidden variables in IST, as labels for state-space trajectories, does not imply any correlation between causally disconnected space-time events.

In this respect, it is worth noting that Bell had an open mind about Superdeterminism, more open at least than the minds of many recent commentators, Sen included. Bell writes:
\begin{quote}
Of course it might be that these reasonable ideas about physical randomisers are just wrong - for the purposes at hand. A theory may appear in which such conspiracies inevitably occur, and these conspiracies may then seem more digestible than the non-localities of other theories. When that theory is announced I will not refuse to listen, either on methodological or other grounds.
\end{quote}
{\sc IST} is a theory where so-called conspiracies are indeed inevitable. However, such so-called conspiracies are not true conspiracies - they \emph{seem} like conspiracies because of an implicit assumption that the distance between a real state and a counterfactual state can be made as small as we like (e.g. in Bell's example, by making a measurement setting depend on the millionth digit of the input to a physical (pseudo-)randomiser. However, as we discussed, the geometrically correct metric for {\sc IST}'s fractal invariant set is the $p$-adic metric. With respect to this metric, the distance between a state on the invariant set and a counterfactual state off the invariant set is at least equal to $p$ and hence large. In this context, the distance between worlds where the parity of the millionth digit of the input to the randomiser are odd and even, is large and the distance is not reduced by making the output of the randomiser dependent on the the billionth or trillionth digit instead of the millionth.

Once again, one should state that Sen's hidden variables are insensitive to the `uncontrollable fluctuations' which could make the difference between the parity of the millionth digit being odd or even. In this sense Sen's hidden variable theory is more like a conventional classical hidden variable theory. {\sc IST}, by contrast, depends on the geometric structure of the invariant set and here these `uncontrollable fluctuations' are critical. Indeed at a formal level, including these `uncontrollable fluctuations' makes {\sc IST} a `non-computable' theory - it is well known now that properties of fractals (does a point in state space lie on the fractal) are generically non-computable \mbox{\cite{Dube:1993}}. By disregarding this, Sen's critique of {\sc IST} is vacuous.

\section{Ontic vs. Epistemic}

Sen concludes that our claim to have a $\psi$-epistemic theory is incorrect, saying
\begin{quote}
``The model is $\psi$-ontic \cite{Harrigan2010OMF,Pusey2012Reality}. This can be noted by the fact that the individual
outcomes depend on the bit-string representation of the quantum state. Given the bit string for a particular run,
the exact quantum state prepared for that run can be inferred.''
\end{quote}

Here we actually sympathise with Sen, since the notion of what is meant by a $\psi$-epistemic theory in the literature is really quite obscure \mbox{\cite{Hance2021Wavefunctions,Hance2021Ensemble}}. In this respect, it is worth returning to basics. We describe uncertainty as epistemic if it is associated with our inability to know everything of relevance. By contrast uncertainty can be said to be ontic if the uncertainty is somehow encoded in the laws of physics. Most quantum physicists believe that the laws of quantum physics incorporate some level of indeterminism, and hence that uncertainty is ontic. Einstein famously believed otherwise. {\sc IST}, as a deterministic theory, can be said to have ontic states, and hence the uncertainty in {\sc IST}, e.g., in knowing which state-space trajectory the real world lies on, is epistemic.

The bit-string is not the ontic state of the model, it is an ensemble of ontic states. Each ``bit'' in the string is an ontic state, labelling a particular trajectory on the invariant set. To give a simple example, a string of the type $(aaaabbbb)$ would correspond to a Hilbert state that is an ensemble of 8 underlying ontic states, the first 4 of which end up in detector eigenstate $|a\rangle$ and the last 4 in eigenstate $|b\rangle$. The first bit of that string could belong to many other Hilbert states, hence the theory is $\psi$-epistemic according to the definition of \cite{Harrigan2010OMF}.

On the other hand, the non-computability of the invariant set suggests that it is not a simple matter of improving the accuracy of our apparatuses to make the unpredictability of quantum physics become predictable. Perhaps we should use the phrase `irreducibly $\psi$-epistemic' to describe this type of uncertainty.

We may note that Sen actually states correctly that the hidden variable $\lambda$ is a function of the position on the string and not the string itself, he just fails to use that definition in his evaluation of the properties of the model.

\section{Misunderstanding of Discreteness}

When relating the past input $\hat{p}$, initial exact orientation $\hat{P}$, new input $\hat{a}$, and final exact orientation $\hat{A}$, Sen states ``We assume for simplicity that $\hat{A}$ can be expressed as the function $\hat{A}=\hat{A}(\hat{P},\hat{p},\hat{a})$.'' He claims having $\hat{A}$ depend on additional variables ``complicates the mathematical analysis without giving any physical insight''. We however dispute this, as having $\hat{A}$ only depend on $\hat{P}$, $\hat{p}$ and $\hat{a}$ neglects the dependence on the discretisation condition.

This is similar to an error he made in an earlier version of his paper, in how he applies the discrete grid implicit to IST over the Bloch sphere. This error led him to claim erroneously that ``The dependence upon the past input $\hat{p}$ arises from the method of rotation which ensures that the final exact orientation $\hat{A}$ has the same orientation relative to $\hat{a}$ as the initial exact orientation $\hat{P}$ had relative to $\hat{p}$,” (or more formally, $\delta \hat{a}=\delta \hat{p}$ in his model, where $\delta \hat{p} = \hat{P}-\hat{p}$, and $\delta \hat{a} = \hat{A}-\hat{a}$). This leads him to assume ``In the single-particle case, the hidden variable $\hat{P}$ and the experimentally set orientations $\hat{p}$ and $\hat{a}$ encoded the exact preparation setting $\hat{A}$ corresponding to the eigenstate $\ket{+}_A$". 

This misunderstands how the discrete grid of valid points (formed by the intersection of the $N$ lines of latitude and $N$ lines of longitude allowed) forces eigenstates measured to the closest point to the eigenstate we would expect for the given operator measured (e.g. eigenstates of $\hat{P}$ instead of those of $\hat{p}$). Depending on the idealised operator measured, the change between the ideal and experimentally-allowed eigenstate has no reason to have to be the same for different operators. Therefore, there is no reason $\delta \hat{A}$ need equal $\delta\hat{P}$, and so there are additional features required to encode the exact preparation setting $\hat{A}$ corresponding to the eigenstate $\ket{+}_A$, than just hidden variable $\hat{P}$ and the experimentally set orientations $\hat{p}$ and $\hat{a}$.

Given this discreteness is a key part of the theory, allowing experimentally-testable differences from standard quantum mechanics to potentially be probed \cite{Hance2021ExpIST}, Sen's misunderstanding of it further undermines the applicability of his model to analysing {\sc IST}.

\section{Conclusion}

We have laid out conceptual mistakes in Sen's recent criticism of Invariant Set Theory. Above all, his hidden-variable model, which somehow ignores the `uncontrollable fluctuations' in the world when considering the meaning of counterfactual measurements, bears no relation to {\sc IST}. This completely nullifies his critiques of {\sc IST} where the same such fluctuations play a critical role in determining the non-computable fractal structure of the invariant set. As we have discussed, with respect to the relevant $p$-adic metric and associated Haar measure of the invariant set, these fluctuations are in no sense small. Hence the normal objections to superdeterminism, of conspiracy and fine tuning, simply do not apply. Sen's hidden-variable model may well be nonlocal, but that has absolutely no bearing whatsoever on the locality or nonlocality of {\sc IST}.

We continue to \emph{insist} that {\sc IST} is deterministic, locally causal and $\psi$-epistemic, and yet reproduces the predictions of quantum mechanics. 

\bigskip

\textit{Acknowledgements}
We thank John Rarity and Sophie Inman for useful conversations. TP is funded by a Royal Society Research Professorship. SH acknowledges support by the Deutsche Forschungsgemeinschaft (DFG, German Research Foundation) under grant number HO 2601/8-1. JRH acknowledges support from the University of York's EPSRC DTP grant EP/R513386/1, and the EPSRC Quantum Communications Hub (funded by the EPSRC grant EP/M013472/1). 

\bibliographystyle{plainurl}
\bibliography{ref.bib}

\begin{thebibliography}{10}

\bibitem{Bell2004Speakable}
John~S Bell.
\newblock {\em Speakable and unspeakable in quantum mechanics: Collected papers
  on quantum philosophy}.
\newblock Cambridge university press, 2 edition, 2004.
\newblock \href {https://doi.org/10.1017/CBO9780511815676}
  {\path{doi:10.1017/CBO9780511815676}}.

\bibitem{Dube:1993}
S.~Dube.
\newblock Undecidable problems in fractal geometry.
\newblock {\em Complex Systems}, 7:423--444, 1993.

\bibitem{Hance2021ExpIST}
Jonte~R Hance, Tim~N Palmer, and John Rarity.
\newblock Experimental tests of invariant set theory.
\newblock {\em arXiv preprint arXiv:2102.07795}, 2021.
\newblock URL: \url{https://arxiv.org/abs/2102.07795}.

\bibitem{Hance2021Wavefunctions}
Jonte~R Hance, John Rarity, and James Ladyman.
\newblock Wavefunctions can simultaneously represent knowledge and reality.
\newblock {\em arXiv preprint arXiv:2101.06436}, 2021.
\newblock URL: \url{https://arxiv.org/abs/2101.06436}.

\bibitem{Hance2021Ensemble}
JR~Hance and S~Hossenfelder.
\newblock The wave-function as a true ensemble.
\newblock {\em arXiv preprint arXiv:2109.02676}, 2021.
\newblock URL: \url{https://arxiv.org/abs/2109.02676}.

\bibitem{Hance2021StatInd}
J.R. Hance, S.~Hossenfelder, and T.N. Palmer.
\newblock Supermeasured: Violating statistical independence without violating
  statistical independence.
\newblock {\em arXiv preprint arXiv:2108.07292}, 2021.
\newblock URL: \url{https://arxiv.org/abs/2108.07292}.

\bibitem{Harrigan2010OMF}
Nicholas Harrigan and Robert~W Spekkens.
\newblock Einstein, incompleteness, and the epistemic view of quantum states.
\newblock {\em Foundations of Physics}, 40(2):125--157, 2010.
\newblock \href {https://doi.org/10.1007/s10701-009-9347-0}
  {\path{doi:10.1007/s10701-009-9347-0}}.

\bibitem{Hossenfelder2020SuperdeterminismGuide}
Sabine Hossenfelder.
\newblock Superdeterminism: A guide for the perplexed.
\newblock {\em arXiv preprint arXiv:2010.01324}, 2020.
\newblock URL: \url{https://arxiv.org/abs/2010.01324}.

\bibitem{Hossenfelder2020Rethinking}
Sabine Hossenfelder and Tim Palmer.
\newblock Rethinking superdeterminism.
\newblock {\em Frontiers in Physics}, 8:139, 2020.
\newblock \href {https://doi.org/10.3389/fphy.2020.00139}
  {\path{doi:10.3389/fphy.2020.00139}}.

\bibitem{Palmer2020Discretization}
TN~Palmer.
\newblock Discretization of the bloch sphere, fractal invariant sets and
  bell’s theorem.
\newblock {\em Proceedings of the Royal Society A}, 476(2236):20190350, 2020.
\newblock \href {https://doi.org/10.1098/rspa.2019.0350}
  {\path{doi:10.1098/rspa.2019.0350}}.

\bibitem{Pusey2012Reality}
Matthew~F Pusey, Jonathan Barrett, and Terry Rudolph.
\newblock On the reality of the quantum state.
\newblock {\em Nature Physics}, 8(6):475--478, 2012.
\newblock \href {https://doi.org/10.1038/NPHYS2309}
  {\path{doi:10.1038/NPHYS2309}}.

\bibitem{Sen2021Analysis}
Indrajit Sen.
\newblock Analysis of the superdeterministic invariant-set theory in a
  hidden-variable setting.
\newblock {\em arXiv preprint arXiv:2107.04761}, 2021.
\newblock URL: \url{https://arxiv.org/abs/2107.04761}.

\end{thebibliography}

\end{document}